\documentclass[aps,prl,twocolumn,superscriptaddress,floatfix]{revtex4-1}

\usepackage{bm}
\usepackage{subfigure}
\usepackage{mathptm}
\usepackage{epsfig}
\usepackage{subfigure}
\usepackage{float}
\usepackage{amsmath,amsfonts,amssymb,graphicx}
\usepackage{pstricks,pst-plot}
\usepackage{graphicx}
\usepackage{dcolumn}

\begin{document}

\title{Chaos in Dirac electron optics: Emergence of a relativistic quantum 
chimera}

\author{Hong-Ya Xu}
\affiliation{School of Electrical, Computer, and Energy Engineering, Arizona State University, Tempe, Arizona 85287-5706, USA}

\author{Guang-Lei Wang}
\affiliation{School of Electrical, Computer, and Energy Engineering, Arizona State University, Tempe, Arizona 85287-5706, USA}

\author{Liang Huang}
\affiliation{School of Physical Science and Technology, and Key Laboratory for Magnetism and Magnetic Materials of MOE, Lanzhou University, Lanzhou, Gansu 730000, China}

\author{Ying-Cheng Lai} \email{Ying-Cheng.Lai@asu.edu}
\affiliation{School of Electrical, Computer, and Energy Engineering, Arizona State University, Tempe, Arizona 85287-5706, USA}

\begin{abstract}

We uncover a remarkable quantum scattering phenomenon in two-dimensional
Dirac material systems where the manifestations of both classically integrable
and chaotic dynamics emerge simultaneously and are electrically controllable.
The distinct relativistic quantum fingerprints associated with different
electron spin states are due to a physical mechanism analogous to chiroptical 
effect in the presence of degeneracy breaking. The phenomenon mimics a 
chimera state in classical complex dynamical systems but here in a relativistic 
quantum setting - henceforth the term ``Dirac quantum chimera,'' associated 
with which are physical phenomena with potentially significant applications 
such as enhancement of spin polarization, unusual coexisting quasibound 
states for distinct spin configurations, and spin selective caustics. 
Experimental observations of these phenomena are possible through, e.g., 
optical realizations of ballistic Dirac fermion systems.   

\end{abstract}

\maketitle

The tremendous development of two-dimensional (2D) Dirac materials such 
as graphene, silicene and germanene~\cite{Novoselovetal:2004,
Novoselovetal:2005,Netoetal:2009,Wehling:2014,Wang2015}, in which the 
low-energy excitations follow the relativistic energy-momentum relation 
and obey the Dirac equation, has led to the emergence of a new area of 
research: Dirac electron optics~\cite{Cse2007,Cheianov2007,Shytov2008,
Beenakker2009,Moghaddam2010,Gu2011,Will2011,Rickhaus2013,Liao2013,Hei2013,
Asm2013,Wu2014,Zhao2015,Peter2015,Leeetal:2015,RPeter2015,Walls2016,
Caridad2016,Gutierrez2016,JLee2016,Chenetal:2016,Settnes2016,Liu2017,
Barnard2017,Jiang2017,Ghahari2017,Zhang2017,Boggild2017}. 
Theoretically, it was articulated early~\cite{Cheianov2007} that Klein 
tunneling and the unique gapless conical dispersion relation can be 
exploited to turn a simply p-n junction into a highly transparent focusing 
lens with a gate-controlled {\em negative refractive index}, producing a 
Vaselago lens for the chiral Dirac fermions in graphene. The negative 
refraction of Dirac fermions obeys the Snell's law in optics and the 
angularly-resolved transmittances in analogy with the Fresnel coefficients 
in optics have been recently confirmed 
experimentally~\cite{Leeetal:2015,Chenetal:2016}. Other works include
various Klein-tunneling junction based electronic counterparts of optical 
phenomena such as Fabry-P\'{e}rot 
resonances~\cite{Shytov2008,Rickhaus2013}, cloaking~\cite{Gu2011,Liao2013}, 
waveguide~\cite{Will2011,Peter2015}, Goos-H\"{a}nchen 
effect~\cite{Beenakker2009}, Talbot effect~\cite{Walls2016}, beam splitter 
and collimation~\cite{RPeter2015,Liu2017,Barnard2017}, 
and even Dirac fermion microscope~\cite{Boggild2017}.
A Dirac material based electrostatic potential junction with a 
closed interface can be effectively tuned to optical guiding and 
acts as an unusual optical dielectric cavity whose effective refractive 
index can be electrically modulated, in which phenomena such as gate 
controlled caustics~\cite{Cse2007}, electron Mie 
scattering~\cite{Hei2013,Caridad2016,Gutierrez2016,JLee2016} and whispering 
gallery modes~\cite{Wu2014,Zhao2015,Jiang2017,Ghahari2017} can arise.
In addition, unconventional electron optical elements have been demonstrated 
such as valley resolved waveguides~\cite{Wu2011,ZMC:2011} and beam 
splitters~\cite{Settnes2016}, electronic birefringent superlens~\cite{Asm2013}
and spin (current) lens~\cite{Moghaddam2010,Zhang2017}. Research on Dirac 
electron optics offers the possibility to control Dirac electron flows in 
a similar way as for light. 

In this Letter, we address the role of chaos in Dirac electron optics.
In nonrelativistic quantum mechanics, the interplay between chaos and 
quantum optics has been studied in microcavity 
lasers~\cite{NSCGC:1996,Nockel1997,GCNNSFSC:1998,Vahala:2003} and deformed 
dielectric microcavities with non-Hermitian physics and wave 
chaos~\cite{Cao2015}. With the development of Dirac electron 
optics~\cite{Cse2007,Cheianov2007,Shytov2008,Beenakker2009,Moghaddam2010,
Gu2011,Will2011,Rickhaus2013,Liao2013,Hei2013,Asm2013,Wu2014,Zhao2015,
Peter2015,Leeetal:2015,RPeter2015,Walls2016,Caridad2016,Gutierrez2016,
JLee2016,Chenetal:2016,Settnes2016,Liu2017,Barnard2017,Jiang2017,
Ghahari2017,Zhang2017,Boggild2017}, the relativistic electronic counterparts
of deformed optical dielectric cavities/resonators have become accessible.
For massless Dirac fermions in ballistic graphene, 
the interplay between classical dynamics and electrostatic confinement
has been studied~\cite{Bardarson2009,Schneider2011,Heinl2013,Schneider2014} 
with the finding that integrable dynamics lead to sharp transport resonances 
due to the emergence of bound states while chaos typically removes the 
resonances. In these works, the uncharged degree of freedom such as electron 
spin, which is fundamental to relativistic quantum systems, was ignored.

Our focus is on the interplay between ray-path defined classical 
dynamics and spin in Dirac electron optical systems. To be concrete, we 
introduce an electrical gate potential defined junction with a ring 
geometry, in analogy to a dielectric annular cavity. Classically, this 
system generates integrable and mixed dynamics with the chaotic fraction of 
the phase space depending on the ring eccentricity and the effective 
refractive index configuration, where the index can be electrically tuned 
to negative values to enable Klein tunneling. Inside the gated region, the 
electron spin degeneracy is lifted through an exchange field from induced 
ferromagnetism, leading to a class of spin-resolved, electrically tunable 
quantum systems of electron optics with massless Dirac fermions (by mimicking 
the photon polarization resolved photonic cavities made from synthesized 
chiral metamaterials). We develop an analytic wavefunction matching 
solution scheme and uncover a striking quantum scattering phenomenon: 
manifestations of classically integrable and chaotic dynamics {\em coexist 
simultaneously} in the system at the same parameter setting, which mimics a 
chimera state in classical complex dynamical systems~\cite{KB:2002,AS:2004,
AS:2006,TNS:2012,HMRHOS:2012,MTFH:2013,YHLZ:2013,YHGL:2015}.
The basic underlying physics is the well-defined, 
spin-resolved, gate-controllable refraction index that dominantly controls 
the ballistic motion of short-wavelength Dirac electrons across the 
junction interface, in which the ray tracing of reflection and refraction 
associated with particles belonging to different spin states generates 
distinct classical dynamics inside the junction/scatterer. Especially, 
with a proper gate potential, the spin-dependent 
refractive index profile can be controlled to generate regular ray dynamics 
for one spin state but generically irregular behavior with chaos for the 
other. A number of highly unusual physical phenomena arise, such 
as enhanced spin polarization with chaos, simultaneous quasiscarred and 
whispering gallery type of resonances, and spin-selective lensing with a 
starkly near-field separation between the local density of states (DOS) 
for spin up and down particles.

Low energy excitations in 2D Dirac materials are described by the Dirac-Weyl 
Hamiltonian $H_0 = v_F\bm{\sigma\cdot p}$, where $v_F$ is the 
Fermi velocity, $\bm{p}=(p_x, p_y)$ is the momentum measured from a 
given Dirac point and $\bm{\sigma}=(\sigma_x, \sigma_y)$ are Pauli 
matrices for sublattice pseudospin. In the presence of a gate potential and 
an exchange field due to the locally induced ferromagnetism inside the whole 
gated region, the effective Hamiltonian is 
$H = v_Fs_0\otimes\bm{\sigma\cdot p} + s_0\otimes\sigma_0\mathcal{V}_{gate}
(\bm{r}) - s_z\otimes\sigma_0\mathcal{M}(\bm{r})$,
where the Pauli matrix $s_z$ acts on the real electron spin space, $s_0$ and 
$\sigma_0$ both are identity matrices, $\mathcal{V}_{gate}(\boldsymbol{r})$ 
and $\mathcal{M}(\boldsymbol{r})$ are the electrostatic and exchange 
potential, respectively. Due to the pseudospin-momentum locking (i.e., 
$\bm{\sigma\cdot p}$), a non-uniform potential couples the two 
pseudospinor components, but the electron spin components are not 
coupled with each other. The exchange field breaks the twofold spin 
degeneracy. Since $[s_z\otimes\sigma_0, H]=0$, the Hamiltonian can be 
simplified as $H_s = H_0 + \mathcal{V}_{gate}(\bm{r})-s\mathcal{M}(\bm{r})$ 
with $s=\pm$ denoting the electron spin quantum number. Because of 
$\mathcal{M}$, the Dirac-type Hamiltonian $H_s$ can give rise to spin 
dependent physical processes.
  
\begin{figure}[h]
\centering
\includegraphics[width=1\linewidth]{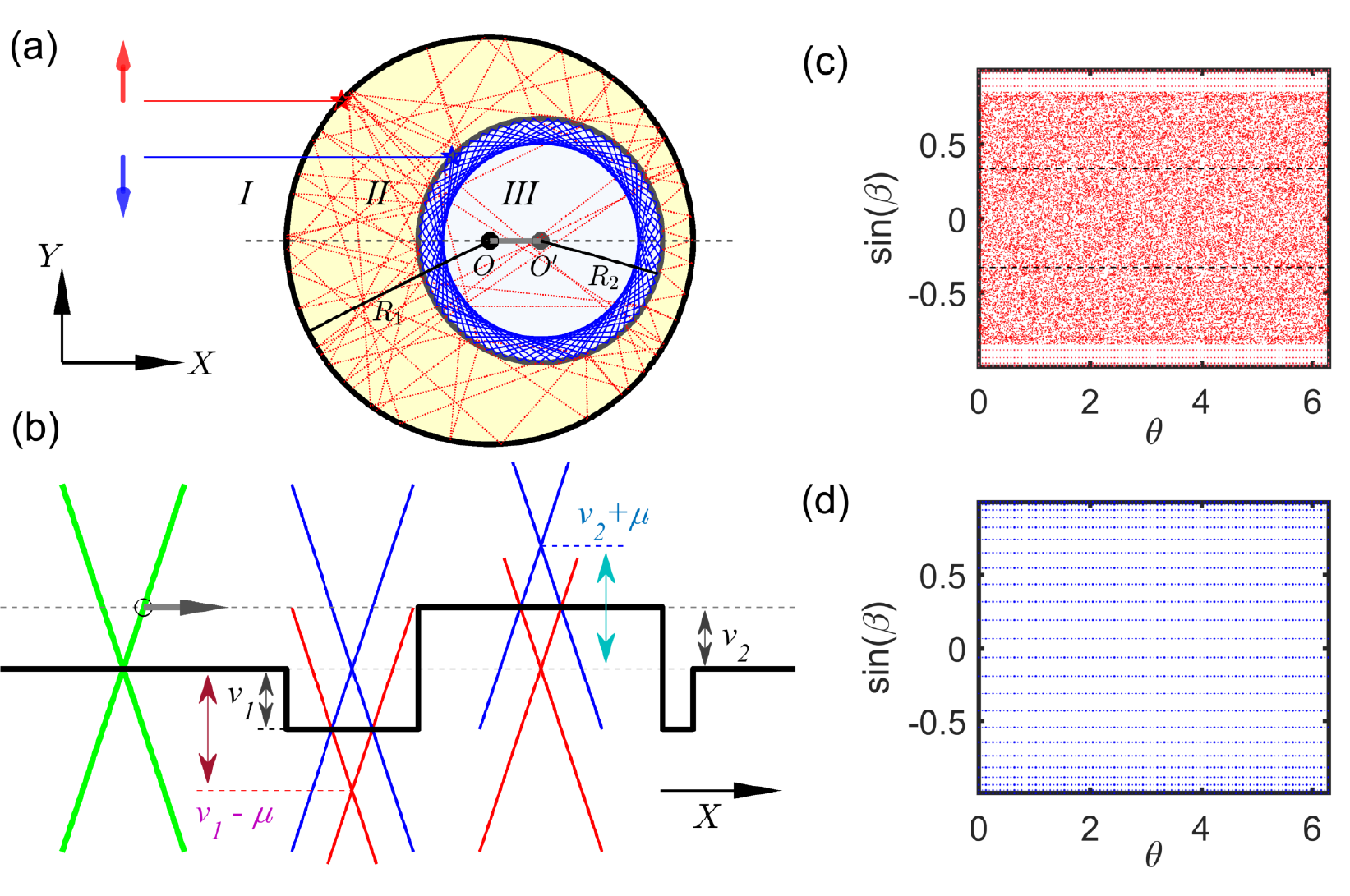}
\caption{ \textbf{Scattering system and classical ray dynamics}.
(a) Annular shaped scattering region 
with eccentricity $\xi = \overline{OO'}$, (b) a 
cross-sectional view, (c,d) chaotic and integrable ray dynamics on the 
Poincar\'{e} surface of section defined by the Birkhoff coordinates 
($\theta$, $\sin\beta$) for spin up and down particles, respectively, where 
$\theta$ denotes the polar angle of a ray's intersection point with the 
cavity boundary and $\beta$ is the angle of incidence with respect to the 
boundary normal. The quantity $\sin\beta$ is proportional to the angular 
momentum and the critical lines for total internal reflection are given by 
$\sin\beta_c=\pm1/n_s$.} 
\label{fig:classical}
\end{figure}

For the ring configuration in Fig.~\ref{fig:classical}(a) and assuming 
the potentials are smooth on the scale of the lattice spacing but 
sharp in comparison with the conducting carriers' wavelength, in 
the polar coordinates $\bm{r}=(r,\theta)$, we have 
$\mathcal{V}_{gate}(\bm{r})=\hbar v_F\nu_1\Theta(R_1 - r)
\Theta(|\bm{r - \xi}|-R_2) + \hbar v_F\nu_2\Theta(R_2 - |\bm{r - \xi}|)$, 
and $\mathcal{M}(\bm{r})=\hbar v_F\mu\Theta(R_1 - r)$, where $\Theta$ 
is the Heaviside step function, $R_2$ is the radius of the small disk gated 
region of strength $\hbar v_F (\nu_2-\nu_1)$ placed inside a larger disk of 
radius $R_1$ ($>R_2$) and strength $\hbar v_F \nu_1$, the displacement vector 
between the disk centers is $\bm{\xi}=(\xi,0)$, and the exchange potential 
has the strength $\hbar v_F \mu$ over the whole gated region. The two circular 
boundaries divide the domain into three distinct regions: $I$: $r >R_1$; 
$II$: $r<R_1$ and $|\bm{r-\xi}|>R_2$; $III$; $|\bm{r-\xi}|<R_2$. For given 
particle energy $E=\hbar v_F\epsilon$, the momenta in the respective regions
are $k_s^{I}=|\epsilon|$, $k_s^{II}=|\epsilon - \nu_1 + s\mu|$, and 
$k_s^{III} = |\epsilon - \nu_2 + s\mu|$. 
Within the gated region, the exchange potential splits the Dirac cone into 
two in the vertical direction in the energy domain while the electrostatic 
potential simply shifts the cone, leading to a spin-resolved, 
gate-controllable annular junction for massless Dirac electrons.

In the short wavelength limit, locally the curved junction interface appears 
straight for the electrons, so the gated regions and the surroundings can be 
treated as optical media. The unusual feature here is that the refractive 
indices are spin-dependent: 
$n_s^{II,III} = (\epsilon +s\mu - \nu_{1,2})/\epsilon$, 
similar to light entering and through a polarization resolved photonic 
crystal~\cite{Gansel2009, ZPLLZ2009}. Given the values of $\epsilon$ and 
$\mu$, depending on the values of $\nu_{1,2}$, the refractive indices for 
the two spin states can be quite distinct with opposite signs. The system is 
thus analogous to a chiral photonic metamaterial based cavity, which 
represents a novel class of Dirac electron optics systems. 

The classical behaviors of Dirac-like particles in the short wavelength 
limit can be assessed using the optical analogy, as done previously for 
circularly curved $p-n$ junctions~\cite{Cse2007,Boggild2017}, where the 
classical trajectories are defined via the principle of least time. 
Because of the spin dependent and piecewise constant nature of the index 
profile, the resulting stationary ray paths for the Dirac electrons are 
spin-resolved and consist of straight line segments. At a junction interface, 
there is ray splitting governed by the spin-resolved Snell's law. On a 
Poincar\'{e} surface of section, the classical dynamics are described by a 
spin-resolved map $F_s$ relating the dynamical variables $\theta$ and $\beta$ 
(Fig.~\ref{fig:classical}) between two successive collisions with the 
interface: $(\theta_i,\sin\beta_i)\mapsto(\theta_{i+1}, \sin\beta_{i+1})$. 
The ray-splitting picture is adequate for uncovering the relativistic 
quantum fingerprints of distinct classical dynamics.

Spin-resolved ray trajectories inside the junction lead to the 
simultaneous coexistence of distinct classical dynamics. For example, for
the parameter setting $\nu_2=-\nu_1=\epsilon=\mu$, i.e., $n_s^{II}=2+s$ and 
$n_s^{III}= s$, for spin up particles ($s=+$), the junction is an eccentric
annular electron cavity characterized by the refractive indices $n_+^{II}=3$
and $n_+^{III}=1$, as exemplified in Fig.~\ref{fig:classical}(b) for 
$\xi=0.3$. However, for spin down particles ($s=-$), the junction appears 
as an off-centered negatively refracted circular cavity with $n_-^{II}=1$ and 
$n_-^{III}=-1$. Figures~\ref{fig:classical}(c) and \ref{fig:classical}(d) show
the corresponding ray dynamics on the Poincar\'{e} surface of section for
spin up and down particles, respectively, where the former exhibit chaos 
while the dynamics associated with the latter are integrable with angular 
momentum being the second constant of motion. 

For a spin unpolarized incident beam, the simultaneous occurrence of 
integrable and chaotic classical dynamics means the coexistence of distinct 
quantum manifestations, leading to the emergence of a Dirac quantum chimera. 
To establish this, we carry out a detailed analysis of the scattering 
matrices for spin-dependent, relativistic quantum scattering and transport 
through the junction. Using insights from analyzing optical dielectric 
cavities~\cite{Hack1997,Hentschel2002} and nonrelativistic quantum billiard 
systems~\cite{Doron1992,Doron1995}, we develop an analytic wave function 
matching scheme at the junction interfaces 
{\color{red}
(See Supplemental Material~\cite{SI} which includes Refs.~\cite{Wigner1955,Smith1960,Schomerus2015,zwillinger2014table,Gutierrez2016,Jiang2017,Weietal:2016})}
to solve the Dirac-Weyl 
equation to obtain the scattering matrix $S$ as a function of the energy 
$E$ as well as the spin polarization $s$ for given system parameters 
$R_2/R_1$, $\xi$, $\nu_{1,2}$ and $\mu$. The Wigner-Smith time 
delay~\cite{Wigner1955,Smith1960} is defined from the $S$-matrix as 
$\tau = -i\hbar\textrm{Tr}\left[S^\dag(\partial S/\partial E)\right]$,
which is proportional to the DOS of the cavity. Large positive values 
of $\tau$ signify resonances associated with the quasibound 
states~\cite{Rotter2017}. Physically, a sharper resonance corresponds to a
longer trapping lifetime and scattering time delay. Previous works 
on wave or quantum chaotic scattering~\cite{BS:1988,BS:1989a,JBS:1990,
MRWHG:1992,LBOG:1992,Ketzmerick:1996,KS:1997,SKGFKDW:1998,KSFN:1999,KS:2000,
HKL:2000,CGM:2000,deMLBAF:2002,CSGFBR:2003,KS:2003,GS:2006,BSS:2010,KFOA:2011,
YHLG:2011a,WYLG:2013} established that classical chaos can smooth out 
(broaden) the sharp resonances and reduce the time delay markedly while 
integrable dynamics can lead to stable, long-lived bound states (or trapping 
modes). 

\begin{figure}[h]
\centering
\includegraphics[width=1\linewidth]{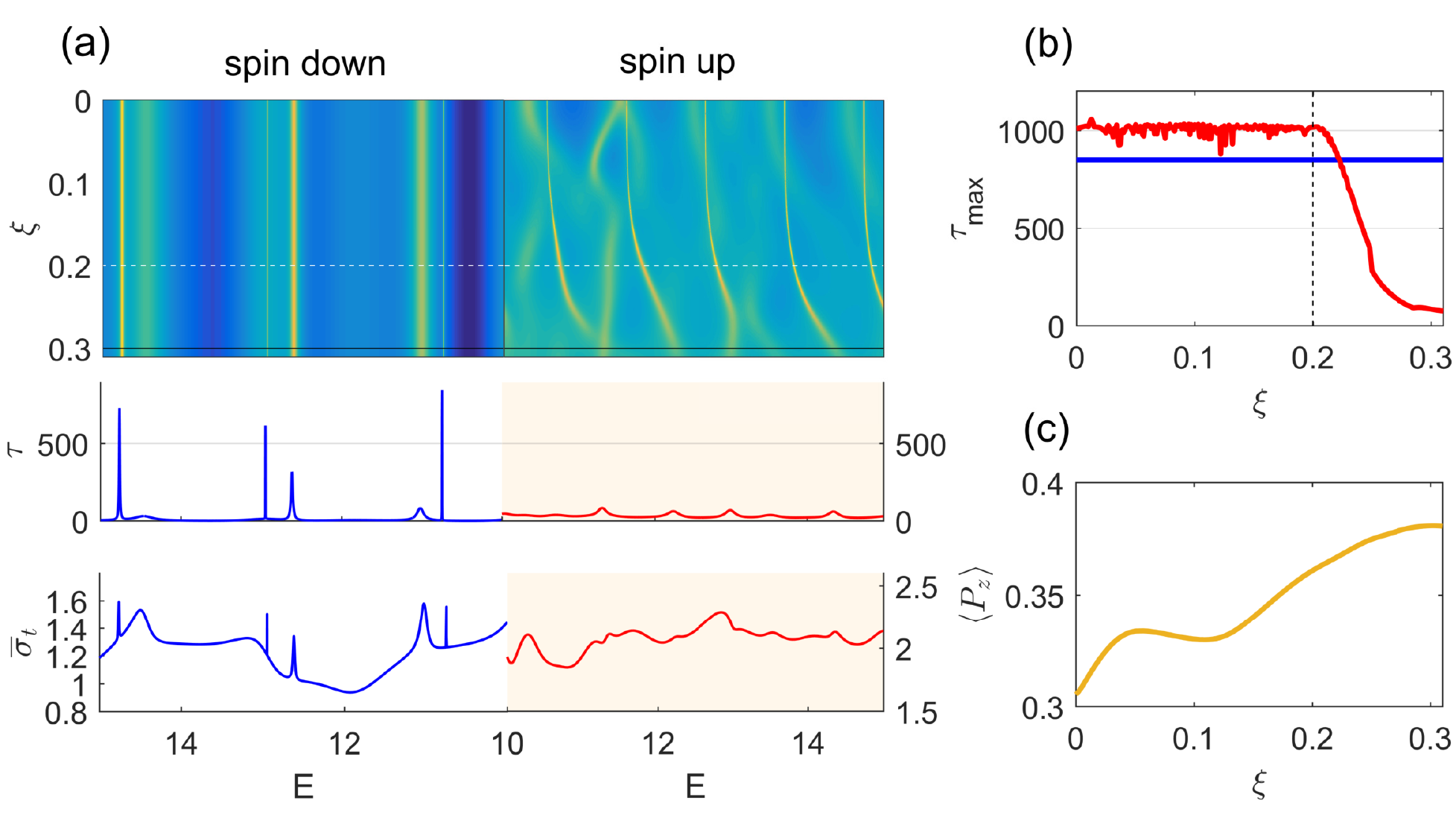}
\caption{ \textbf{A Dirac quantum chimera}. (a) top: Contour map of 
dimensionless Wigner-Smith time delay (on a logarithmic scale) versus energy 
$E$ and eccentricity $\xi$ for spin down (left) and up (right) cases, where 
the bright yellow color indicates larger values. Middle and bottom panels: 
time delay and total cross section averaged over all directions of the 
incident waves versus $E$, respectively, for $\xi=0.3$. (b) Dependence of 
the maximum time delay on $\xi$ (red: spin up; blue: spin-down). (c) Energy 
averaged spin polarization versus $\xi$.}
\label{fig:DQC}
\end{figure}

We present concrete evidence for Dirac quantum chimera. 
Figure~\ref{fig:DQC}(a) shows, for $R_2/R_1=0.6$, $\mu=-\nu_1=5$ and 
$\nu_2 = 45$, the dimensionless time delay (on a logarithmic scale) 
versus the eccentricity $\xi$ and energy $E$ (in units of $\hbar v_F/R_1$). 
Figure~\ref{fig:DQC}(b) shows the maximum time delay [within the given 
energy range in Fig.~\ref{fig:DQC}(a)] versus $\xi$ for spin-up (red) 
and spin-down (blue) particles. There are drastic 
changes in the time delay as the energy is varied, which are characteristic 
of well-isolated, narrow resonances and imply the existence of relatively 
long-lived confined modes. There is a key difference in the resonances 
associated with the spin up and down states: the former depend on the 
eccentricity parameter $\xi$ and are greatly suppressed for $\xi>0.2$, 
while the latter are independent of $\xi$. For example, the middle panel
of Fig.~\ref{fig:DQC}(a) shows that, for a severely deformed structure 
($\xi = 0.3$), there are sharp resonances with high peak values of the 
time delay for the spin down state, but none for the spin up state. The 
suppression of resonances associated with the spin up state is consistent 
with the behavior of the total cross section $\overline{\sigma}_t$
(averaged over the directions of the incident wave) given in terms of the 
$S$-matrix elements by 
$\overline{\sigma}_t = (2k)^{-1}\sum_{m,l=-\infty}^{\infty}
\left|S_{ml}-\delta_{ml}\right|^2$,
as shown in the bottom panel of Fig.~\ref{fig:DQC}(a). Because the classical 
dynamics for massless fermions in the spin up and down states are chaotic and 
integrable, respectively [c.f., Figs.~\ref{fig:classical}(c,d)], there is
{\em simultaneous} occurrence of two characteristically different quantum 
scattering behaviors for a spin unpolarized beam: one without and another 
with \emph{sharp} resonances. This striking contrast signifies a Dirac 
quantum chimera. 

Are there unexpected, counterintuitive physical phenomena associated with
a Dirac quantum chimera? Yes, there are! Here we present two and point out 
their applied values.

\begin{figure}[h]
\centering
\includegraphics[width=1\linewidth]{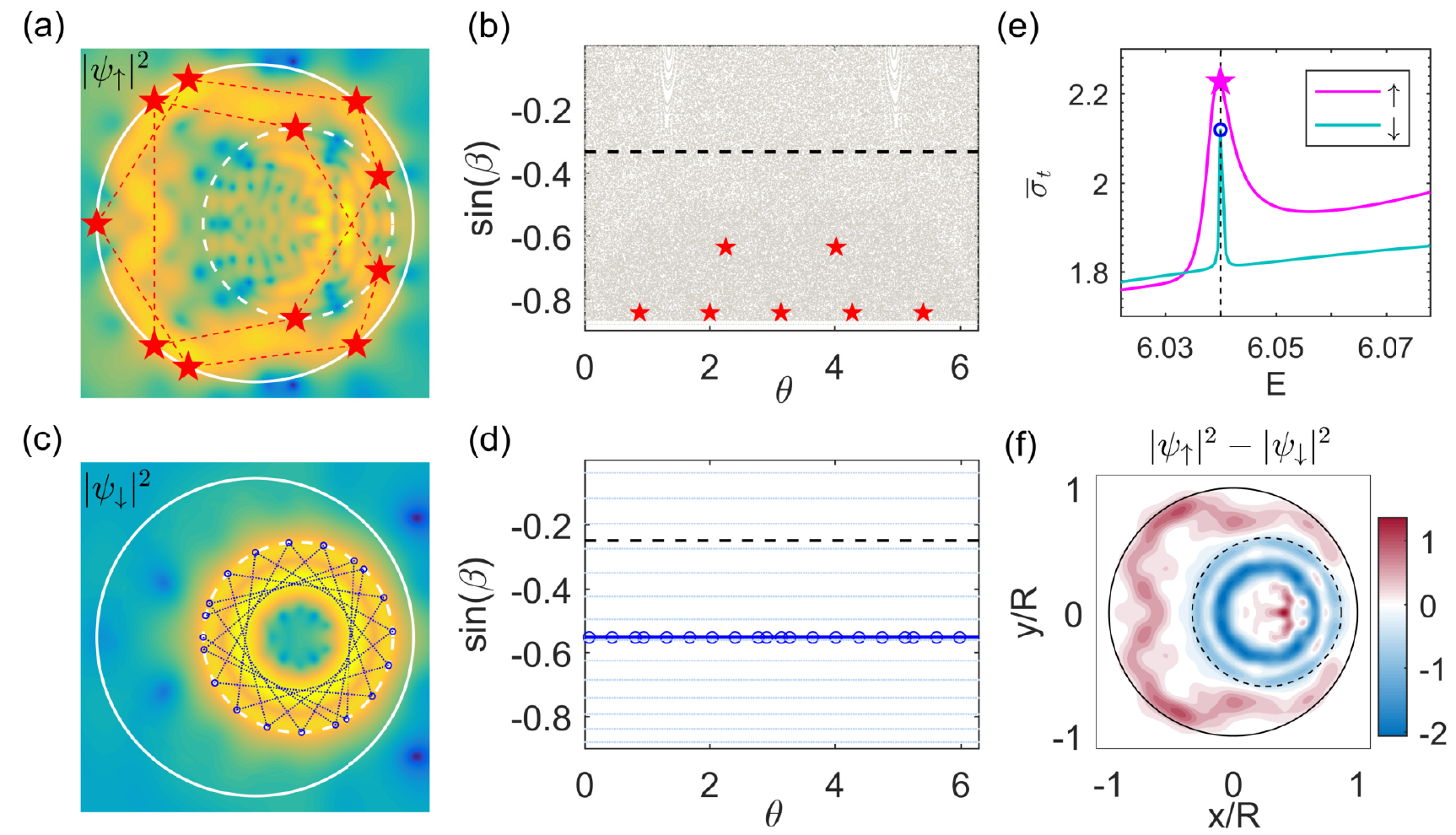}
\caption{ \textbf{Spin polarized scarred and regular whispering-gallery-mode 
resonances as a result of Dirac quantum chimera}. (a,c) Real space 
probability densities (on a logarithmic scale) of the representative 
quasibound states for spin-up and spin-down Dirac electrons, respectively. 
For the spin-up particles, the spinor wave solution is scarred by an unstable 
periodic ray trajectory obeying the Snell's law, as indicated by the red 
dashed path with highlighted pentagram markers. The spin-down Dirac electrons 
are associated with a whispering gallery ray path due to the continuous total 
internal reflections denoted by the blue dotted segments. (b,d) The 
corresponding phase-space representations with regions below the critical 
black dash lines satisfying the total internal reflection at the boundary.
The distinct quasibound modes are from simultaneous resonances under the 
same system parameters, leading to a relativistic quantum chimera. Further 
signatures of the chimera state can be seen in the plot of the total cross 
section versus the particle energy for different spin states (e) and a net 
spin distribution with a dramatic spin-resolved separation in the real space 
confined inside the cavity (f).}
\label{fig:scar-wgm}
\end{figure}

The first is spin polarization enhancement, which has potential 
applications to Dirac material based spintronics. A general way to define spin 
polarization is through the spin conductivities $G^{\downarrow(\uparrow)}$ as 
$P_z = (G^{\downarrow}-G^{\uparrow})/(G^{\downarrow}+G^{\uparrow})$.
Imagine a system consisting of a set of sparse, randomly distributed, 
identical junction-type of annular scatterers, and assume that the scatterer
concentration is sufficiently low ($n_c\ll1/R_1^2$) so that multiple 
scattering events can be neglected. In this case, the spin conductivities 
can be related to the transport cross section as 
$G^{\downarrow(\uparrow)}/G_0=k/(n_c\sigma_{tr}^{\downarrow(\uparrow)})$, where 
$G_0$ is the conductance quantum and $\sigma_{tr}^{\downarrow(\uparrow)}$ 
can be calculated from the $S$-matrix. For a spin unpolarized incident 
beam along the $x$-axis with equal spin up and down populations, we calculate 
the average spin polarization over a reasonable Fermi energy range 
as a function of the eccentricity $\xi$, as shown in Fig.~\ref{fig:DQC}(c). 
For $\xi>0.2$ so classical chaos is relatively well developed and 
a Dirac quantum chimera emerges, there is robust enhancement of spin 
polarization. From the standpoint of classical dynamics, the scattering 
angle is much more widely distributed for spin up particles (due to chaos) 
as compared with the angle distribution for spin down particles with 
integrable dynamics, leading to a larger effective resistance for spin up 
particles. From an applied perspective, the enhancement of spin 
polarization brought about by a Dirac quantum chimera can be exploited for 
developing spin rheostats or filters, where one of the spin resistances, 
e.g., $R^{\uparrow}\propto 1/G^{\uparrow}$, can be effectively modulated 
through tuning the deformation parameter $\xi$ so as to induce classically 
chaotic motion for one type of polarization but integrable dynamics for 
another.  

\begin{figure}[h]
\centering
\includegraphics[width=1\linewidth]{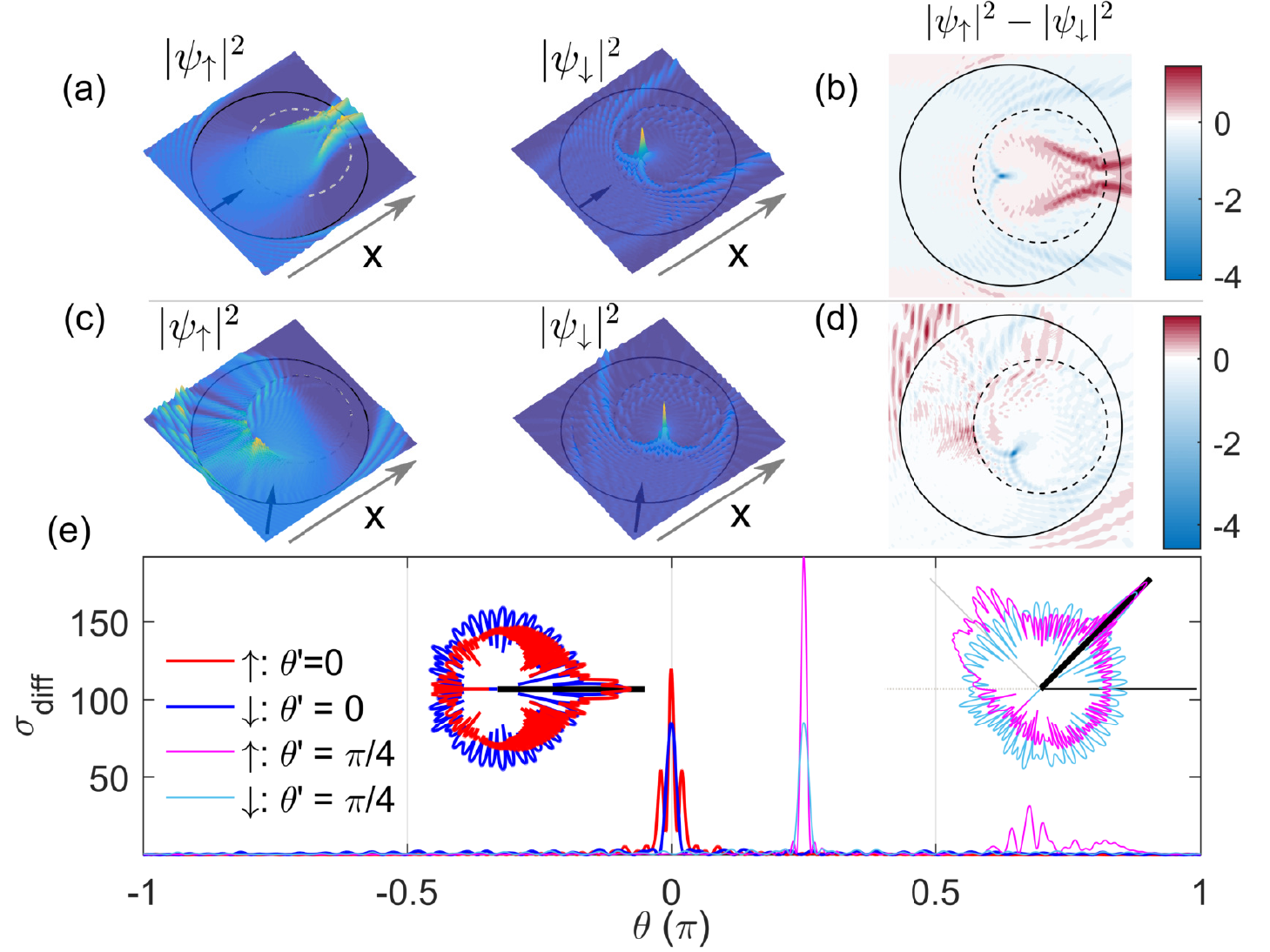}
\caption{ \textbf{Spin-selective caustic lens and skew scattering associated 
with a Dirac quantum chimera}. (a) Caustic patterns resulting from the 
scattering of a spin unpolarized planar incident wave traveling along the 
positive $x$-axis ($\theta'=0$) with relatively short wavelength, i.e., 
$kR_1=70\gg1$, and (c) from scattering of the wave propagating along the 
direction that makes an angle $\theta'=\pi/4$ with the $x$ axis. (b,d) The 
corresponding spatially resolved near field net spin distributions measured 
by the difference $|\psi_\uparrow|^2-|\psi_\downarrow|^2$, respectively. 
(e) The resulting far-field behavior characterized by the angular distributions
of spin-dependent differential cross sections with symmetric profiles for 
$\theta'=0$ (left inset) and spin-selective asymmetric one for 
$\theta'=\pi/4$ (right inset), where both insets are plotted by the eighth 
root of $\sigma_{diff}^{\uparrow(\downarrow)}$  in order to weaken the 
drastic contrast variation in magnitude for better visualization. Parameters 
are $\xi=0.27$, $R_2/R_1=0.6$, $\nu_2=\mu=-\nu_1=70$ and $E=70$.}
\label{fig:Lensing}
\end{figure}

The second phenomenon is resonance and lensing associated with a Dirac 
quantum chimera. Figures~\ref{fig:scar-wgm}(a-f) show, for $\xi=0.27$ 
(in units of $R_1$), $R_2/R_1=0.6$, $\nu_2=4\nu_1=-4\mu=24.16$ (in units 
of $1/R_1$) and $E=6.04$ (in units of $\hbar v_F/R_1$), a resonant 
(quasibound) state, in which the spatially separated, spin resolved local 
DOS is confined inside the cavity. The spin up state is concentrated about 
a particular unstable periodic orbit without the rotational symmetry 
[Figs.~\ref{fig:scar-wgm}(a) and \ref{fig:scar-wgm}(b)] and exhibits a 
scarring pattern with a relatively short lifetime characterized by a wider 
resonance profile, as shown in Fig.~\ref{fig:scar-wgm}(e). Spin down particles 
are trapped inside the inner disk by a regular long-lived whispering gallery 
mode associated with the integrable dynamics [Figs.~\ref{fig:scar-wgm}(c) 
and \ref{fig:scar-wgm}(d)]. The Dirac quantum chimera thus manifests itself 
as the simultaneous occurrence of a magnetic scarred quasibound state and 
a whispering gallery mode excited by an incident wave 
with equal populations of spin up and down particles, as shown in 
Fig.~\ref{fig:scar-wgm}(f), a color-coded spatial distribution of the 
difference between the local DOS for spin up and down particles. 

In the sufficiently short wavelength regime where the ray picture becomes
accurate, a spin resolved lensing behavior arises, due to the simultaneous 
occurrence of two distinct quantum states associated with the chimera state. 
The cavity can be regarded as an effective electronic Veselago lens 
with a robust caustic function for spin down particles but the spin up 
particles encounter simply a conventional lens of an irregular shape. 
Particularly, for a spin-unpolarized, planar incident wave, a spin-selective 
caustic behavior arises, as shown in Figs.~\ref{fig:Lensing}(a-d) through 
the color-coded near-field patterns. There is a pronounced lensing caustic 
of the cusp type for the spin down state while a qualitatively distinct 
lensing pattern occurs for the spin up state. A consistent far-field angular 
distribution of the differential cross section is shown in 
Fig.~\ref{fig:Lensing}(e), which gives rise to well-oriented/collimated, 
spin-dependent far-field scattering with the angle resolved profile shrinked 
into a small range 
due to the lensing effect. Despite lack of robust lensing, the spin up 
particles in general undergo asymmetric scattering, which can lead to 
spin-polarized transverse transport in addition to longitudinal spin 
filtering. 

To summarize, we uncover a Dirac quantum chimera - a type of relativistic 
quantum scattering states characterized by the simultaneous coexistence of 
two distinct types of behaviors as the manifestations of classical chaotic 
and integrable dynamics, respectively. The physical origin of the chimera state 
is the optical-like behavior of massless Dirac fermions with both spin and 
pseudospin degrees of freedom, which together define a spin-resolved Snell's 
law governing the chiral particles' ballistic motion. The phenomenon is 
predicted analytically based on quantum scattering from a gate-defined 
annular junction structure. The chimera has striking physical consequences 
such as spin polarization enhancement, unusual quantum resonances, and 
spin-selective lensing, which are potentially exploitable for developing 
2D Dirac material based electronic and spintronic devices. 


This work is supported by ONR under Grant No.~N00014-16-1-2828. 
L.H. is also supported by NNSF of China under Grant No.~11775101.


%
\end{document}